\documentclass[aps,amsmath,amssymb,prd,preprintnumbers,superscriptaddress,nofootinbib]{revtex4}

\usepackage{amsmath}
\usepackage{bm}
\usepackage[dvips]{graphicx}
\usepackage{color}
\newcommand{\be}{\begin{equation}}

\newcommand{\ee}{\end{equation}}
\newcommand{\bea}{\begin{eqnarray}}
\newcommand{\eea}{\end{eqnarray}}

\begin{document}
\preprint{BARI-TH 567/07}

\author{R.~Anglani}\email{roberto.anglani@ba.infn.it}
\affiliation{Institute for Theoretical Physics, University of Wroclaw, Max Born place 9,
50204 Wroclaw, Poland}\affiliation{Dipartimento di Fisica,
Universit\`a di Bari, I-70126 Bari, Italia}\affiliation{I.N.F.N.,
Sezione di Bari, I-70126 Bari, Italia}

\author{R.~Gatto}\email{raoul.gatto@physics.unige.ch}
\affiliation{D\'epart. de Physique Th\'eorique, Universit\'e de
Gen\`eve, CH-1211 Gen\`eve 4, Suisse}

\author{N.~D.~Ippolito}\email{nicola.ippolito@ba.infn.it}
\affiliation{Dipartimento di Fisica, Universit\`a di Bari, I-70126
Bari, Italia}\affiliation{I.N.F.N., Sezione di Bari, I-70126 Bari,
Italia}

\author{G.~Nardulli}\email{giuseppe.nardulli@ba.infn.it}

\author{M.~Ruggieri}\email{marco.ruggieri@ba.infn.it}
\affiliation{Dipartimento di Fisica, Universit\`a di Bari, I-70126
Bari, Italia}\affiliation{I.N.F.N., Sezione di Bari, I-70126 Bari,
Italia}

\date{\today}

\title{Superfluid and Pseudo-Goldstone Modes
in Three Flavor Crystalline Color Superconductivity}

\begin{abstract}

We study the bosonic excitations in the favorite cubic three flavor
crystalline LOFF phases of QCD.  We calculate in the Ginzburg-Landau
approximation the masses of the eight pseudo Nambu-Goldstone Bosons
(NGB) present in the low energy theory. We also compute the decay
constants of the massless NGB Goldstones associated to superfluidity
as well as those of the eight pseudo NGB. Differently from the
corresponding situation in the Color-Flavor-Locking phase, we find
that meson condensation phases are not expected in the present
scenario.
\end{abstract}

\maketitle

\section{Introduction}
The last few years  have witnessed a conspicuous research activity
on Quantum Chromodynamics (QCD) in extreme conditions of baryon
density and/or temperature. Though mostly performed by model
calculations, these investigations suggest a rich structure for the
QCD phase diagram in the high density and low temperature
regime~\cite{Abuki:2004zk,Ruster:2005jc,Blaschke:2005uj,Hatsuda:2006ps,Yamamoto:2007ah}.
As a matter of fact, in these conditions a phase transition from
hadrons to deconfined quark matter is expected to
occur~\cite{Collins:1974ky}. Because of the attractive interaction
in the antisymmetric color channel the ground state of deconfined
quark matter is reorganized to be a color
superconductor~\cite{Rapp:1997zu,Alford:1997zt} (for reviews on
color superconductivity
see~\cite{Rajagopal:2000wf,Alford:2001dt,Nardulli:2002ma,Buballa:2003qv,
Rischke:2003mt,Huang:2004ik}).

It is well accepted that at asymptotic densities the ground state of
three flavor quark matter is the color-flavor-locked (CFL)
phase~\cite{Alford:1998mk}. In this exotic state of matter both
color and flavor symmetries are spontaneously broken because of the
non-vanishing of the expectation value of a diquark operator.
Nevertheless, a residual subgroup linking color and flavor is left
unbroken. At lower densities, as probably important for the cores of
neutron stars, less symmetric pairing patterns have to be
considered. This is so because effects of the quark masses, as well
as of electrical and color neutrality conditions, cause a mismatch
of the Fermi surfaces of the pairing quarks.  Examples include spin
one
pairing~\cite{Schafer:2000tw,Alford:2002rz,Schmitt:2004et,Aguilera:2005tg,Marhauser:2006hy},
homogeneous gapless two flavor~\cite{Shovkovy:2003uu} and three
flavor~\cite{Alford:2003fq} superconductivity, and crystalline color
superconductivity with two~\cite{Alford:2000ze,Casalbuoni:2004wm}
and three
flavors~\cite{Casalbuoni:2005zp,Mannarelli:2006fy,Casalbuoni:2006zs,Ippolito:2007uz,Rajagopal:2006ig}.
Crystalline superconductors are known as
Larkin-Ovchinikov-Fulde-Ferrell (LOFF) states, from studies in
condensed matter superconductors with magnetic
impurities~\cite{LOFF}. Homogeneous gapless phases are affected by
chromo-magnetic instability~\cite{Huang:2004bg,Casalbuoni:2004tb}.
This means that the screening masses of some of the gluons are
imaginary. On the other hand, it has been shown that this
instability can be cured either by the crystalline color
superconductivity~\cite{Giannakis:2004pf,Ciminale:2006sm,Gorbar:2005tx,Gatto:2007ja}
or by gluon condensed phases~\cite{Gorbar:2007vx}, as well as
condensed meson current states~\cite{Huang:2005pv}.

It has been found in Refs.~\cite{Rajagopal:2006ig,Ippolito:2007uz}
that there exists a window of values of the baryon chemical
potential $\mu$  where the three flavor crystalline color
superconductor is the most favorable candidate to represent the
ground state of high density QCD.  This fact, together with the
 possibility of the crystalline
phases in the core of a compact star, makes interesting the study of
the quantum excitations of the ground state. This is the aim of our
work. In particular we study the quantum excitations (Goldstone or
pseudo-Goldstone modes) in the three flavor LOFF phase of QCD,
arising from the spontaneous breaking of the global symmetries. The
spontaneous breaking of $SU(3)_A$ implies the existence of eight
pseudo-Goldstone modes, while that of $U(1)_V$ (superfluid mode)
entails a massless Goldstone mode. The superfluid mode is massless
even in presence of massive quarks and/or differences of the quark
chemical potentials. On the other hand, finite quark masses and
chemical potential differences cause non-vanishing masses for the
$SU(3)_A$ pseudo-Goldstone modes. In the CFL case it has been found
that even if one takes $m_u = m_d =0$ and $m_s \neq 0$, a finite
mass is found for the excitations with the quantum numbers of the
kaons~\cite{Beane:2000ms,Bedaque:2001je,Schafer:2001za,Son:1999cm,Manuel:2000wm,
Buballa:2004sx,Werth:2006tj,Forbes:2004ww,Ruggieri:2007pi}. Since,
in the case $m_u = m_d =0$, the squared masses of these excitations
are found to be negative, a kaon condensation may occur. Therefore
we are led to investigate the possibility of meson condensation in
the three flavor LOFF phase as well. To do that we compute in this
paper the masses of the pseudo Goldstone modes. Our main result is
the absence of meson condensation and the stability of the LOFF
phase at least to the order $\Delta^2/\delta\mu^2$.

The plan of the paper is as follows. In Section~\ref{sec:quarks} we
derive the effective quark lagrangian in the crystalline LOFF phase.
In Section~\ref{sec:counting} we discuss the coupling of the quarks
to the Goldstones and we  derive the effective action of the scalar
excitations and their parameters. Finally, in Section~\ref{concl} we
draw our conclusions and discuss possible prosecutions of this work.

\section{The effective quark lagrangian}\label{sec:quarks}
In this paper we deal with three flavor quark matter, whose
interaction is modeled  by a local Nambu-Jona Lasinio (NJL)
lagrangian, evaluated in the high density effective theory as in
Ref.~\cite{Casalbuoni:2005zp} (see below Eq.~\eqref{eq:Lagr1}). At
finite chemical potential and in presence of color condensation the
quark lagrangian is given by
\begin{equation}
{\cal L} = \bar\psi\left(i\partial_\mu \gamma^\mu +
\hat\mu\gamma_0\right)\psi - M_f \bar{\psi}_f \psi_f +{\cal
L}_\Delta~.\label{eq:lagr1MU}
\end{equation}
In the above equation $\hat\mu$ is the quark chemical potential
matrix, with color and flavor indices. It depends on $\mu$ (the
average quark chemical potential), $\mu_e$ (the electron chemical
potential), and $\mu_3,\,\mu_8$ (color chemical potentials)
\cite{Alford:2003fq}. To implement color and electric neutrality it
is sufficient to consider only these chemical potentials,
 related to the charge matrix $Q$ and to
 the diagonal color operators $T_3
= \frac 1 2 {\rm diag}(1,-1,0)$ and $T_8 = \frac{1}{2 \sqrt 3 }{\rm
diag}(1,1,-2)$. In general one should introduce a color chemical
potential for each $SU(3)$ color charge; however, as shown
in~\cite{Buballa:2005bv}, for the condensate with the color-flavor
structure considered in this paper it is enough to consider only
$\mu_3$ and $\mu_8$, since the charges related to the other color
generators automatically vanish. Therefore the matrix $\hat\mu$ is
written as follows\be {\hat\mu}_{ij}^{\alpha\beta} = \left(\mu
\delta_{ij} - \mu_e Q_{ij}\right)\delta^{\alpha\beta}+ \delta_{ij}
\left(\mu_3 T_3^{\alpha\beta}+\frac{2}{\sqrt 3}\mu_8
T_8^{\alpha\beta}\right) \label{9}\ee with $Q= {\rm diag}
(2/3,-1/3,-1/3)$ ($i,j =1,3 $ flavor indices; $\alpha,\beta =1,3 $
colour indices).

The term ${\cal L}_\Delta$ is responsible for color condensation,
and in the mean field approximation it is given
by~\cite{Rajagopal:2006ig,Ippolito:2007uz}
\begin{equation}
{\cal L}_\Delta = -\frac{1}{2}\sum_{I=1}^3\left(\Delta_{I}({\bm
r})\psi_{i\alpha }^\dagger \gamma_5\epsilon^{\alpha\beta
I}\epsilon_{i j I} C \psi_{\beta j}^* + h.c. \right) -
\frac{\Delta_I({\bm r})\Delta_I^*({\bm
r})}{3G}~.\label{eq:LagrDelta2}
\end{equation}
Eq.~\eqref{eq:LagrDelta2} describes the fact that in the ground
state one has a non-vanishing expectation value of the diquark field
operator
\begin{equation}
\langle\psi({\bm r})_{i\alpha} C\gamma_5 \psi({\bm r})_{\beta
j}\rangle \propto \Delta_I({\bm r}) \epsilon_{\alpha \beta
I}\epsilon_{ijI} \neq 0~.\label{eq:SSB}
\end{equation}As discussed in the Introduction we will consider
kinematical conditions favoring the LOFF phase
\cite{Casalbuoni:2005zp,Mannarelli:2006fy,Rajagopal:2006ig}. In this
case the ${\bm r}$-dependence of the gap parameters is given by a
linear combination of plane waves,
\begin{equation}
\Delta_I({\bm r}) = \Delta_I \sum_{a=1}^{P_I}e^{2i{\bm q}_I^a \cdot
{\bm r}}~.
\end{equation}

In~\cite{Rajagopal:2006ig} several crystalline structures have been
considered, and their free energy has been computed in the
Ginzburg-Landau approximation. All the structures considered
in~\cite{Rajagopal:2006ig} have $\Delta_1=0$, $\Delta_2 = \Delta_3
\equiv\Delta$, $P_2 = P_3 \equiv P$. It was found that a crystalline
color superconductive phase exists in the following interval:
\begin{equation}
2.88\Delta_0 \le \frac{M_s^2}{\mu} \le
10.36\Delta_0~,\label{eq:window2}
\end{equation}
where $\Delta_0$ is the CFL gap in the chiral limit and $M_s$ is the
in-medium strange quark mass. In more detail, for $2.88\Delta_0 \le
M_s^2/\mu \le 6.20\Delta_0$ the ground state of three flavor quark
matter is the CubeX. In this structure $P = 4$; for each pairing
channel the wave vectors $\{{\bm q}_I\}$ form a square, and the two
squares are arranged in such a way that they point to the vertices
of a cube. In the remaining region the favored structure is the
2Cube45z in which $P= 8$; each wave vector set $\{{\bm q}_I\}$ forms
a cube, and the two cubes are rotated by 45 degrees around an axes
perpendicular to one of the faces of the cube. In the following we
shall concentrate on these two crystalline structures.

Finally $M_f$ denote the in-medium quark mass of the flavor $f$. In
the crystalline superconductive phases the in-medium quark masses
have been evaluated self-consistently in~\cite{Ippolito:2007uz}. It
was found that for values of $\mu$ high enough for the condensation
in the three flavor case to occur, the constituent $u$ and $d$ quark
masses numerically coincide with their bare values $M_u\sim m_u$,
$M_d \sim m_d$. In Ref.~\cite{Ippolito:2007uz} it was found the LOFF
window
\begin{equation}
442\,~\text{ MeV}\,<\,\mu\,<\,515~\text{ MeV} ~,\label{cubo}
\end{equation}
and correspondingly $M_s$ belongs to the interval $270-463$ MeV.

In this paper we adopt the high density effective description of
QCD~\cite{Hong:1998tn,Nardulli:2002ma,Schafer:2003jn}. This
approximation amounts to consider only the quarks with momenta close
to the Fermi surface and it is justified since in the weak coupling
regime we are interested in here the quarks living in the depth of
the Fermi sphere are Pauli blocked and irrelevant for the dynamics.
Furthermore the antiparticle poles can be neglected in the quark
propagator, as they give rise to operators that are formally
suppressed by inverse powers of $\mu$.

The high density effective lagrangian of the quarks in the three
flavor LOFF phase of QCD, derived from Eq.~\eqref{eq:lagr1MU}, is
obtained in Ref.~\cite{Casalbuoni:2005zp}; therefore here we simply
quote the result in the momentum space, namely
\begin{equation}
{\cal L}=\frac{1}{2}\int\!\frac{d{\bm n}}{4\pi}~\chi^\dagger_A
\left(\begin{array}{cc}
        V\cdot\ell~\delta_{AB} + \delta\mu_{AB} & -\Delta_{AB} \\
        -\Delta^\star_{AB} & \tilde{V}\cdot\ell~\delta_{AB} - \delta\mu_{AB}
      \end{array}
\right)\chi_B ~ + L\rightarrow R~.\label{eq:Lagr1}
\end{equation}
Here $A=1,\dots,9$ is a color-flavor index; the rotation to the new
basis is performed by means of the matrices $F_A$ defined
in~\cite{Casalbuoni:2004tb}. The quark momenta are measured as ${\bm
p} = \mu{\bm n}+{\bm \ell}$, $p_0 = \ell_0$, with ${\bm n}$ a unit
vector denoting the Fermi velocity of the quarks and $\mu$ is a
reference large momentum (usually one takes $\mu$ equal to the
baryon chemical potential, but in the LOFF phase it is more
convenient to measure the momenta with respect to the $u$ Fermi
momentum, see below). The chemical potential of the quark with index
$A$ is written as $\mu_A = \mu + \delta\mu_A$ and $\delta\mu_{AB}
\equiv \delta\mu_A \delta_{AB}$. In the three flavor LOFF phase one
can assume $\mu_3 = \mu_8 = 0$ and $\mu_e =
M_s^2/4\mu$~\cite{Casalbuoni:2006zs}, therefore the quark chemical
potential matrix can be written as $[\text{diag}(\mu_u,\mu_u +
2\delta\mu,\mu_u -2\delta\mu)]_{ij}\otimes\delta_{\alpha\beta}$ with
\be \delta\mu = \mu_e/2 = M_s^2/8\mu\ .\ee

The gap matrix is given by $\Delta_{AB} = \Delta_I({\bm
r})\text{Tr}[\epsilon_I F_A^T \epsilon_I F_B]$; the explicit form is
in \cite{Casalbuoni:2004tb}. Finally, we have introduced the
Nambu-Gorkov doublet
\begin{equation}
\chi = \left(\begin{array}{c}
               \psi({\bm n}) \\
               C\psi^*(-{\bm n})
             \end{array}
\right)~.
\end{equation}
Here $\psi({\bm n})$ is a positive energy field with velocity ${\bm
n}$; the projection is achieved by the projectors $P_\pm = (1 \pm
{\bm \gamma_0 {\bm\gamma}}\cdot{\bm n})/2$.

\section{Goldstone modes}\label{sec:counting}
The symmetry group of three massless flavor QCD is
\begin{equation}
\rm G = SU(3)_c\otimes SU(3)_V \otimes SU(3)_A \otimes U(1)_V
\otimes U(1)_A~. \label{eq:Gqcd}
\end{equation}
Here and in the following we assume that the baryon chemical
potential is high enough to restore the $U_A(1)$ symmetry. In the
neutral unpaired quark matter with a massive strange quark the
flavor symmetries of G are also explicitly broken by mass terms.
Even with vanishing $M_u$ and $M_d$
 the chemical potential matrix
$[\text{diag}(\mu_u,\mu_u + 2\delta\mu,\mu_u
-2\delta\mu)]_{ij}\otimes\delta_{\alpha\beta}$ ($\delta\mu = \mu_e/2
= M_s^2/8\mu$)  is invariant only under the transformations of
$SU(3)_V$ and $SU(3)_A$ generated by $\lambda_3$ and $\lambda_8$. In
the sequel we will keep track not only of $M_s\neq 0$, but also of
the small corrections due to $M_u$ and $M_d$.

In the CFL case~\cite{Alford:1998mk} the condensates leave unbroken
the global symmetry group $SU(3)_{c+L+R}$, which is explicitly
broken in the present case by mass terms and chemical potential
differences. On the other hand the color gauge group is
spontaneously broken, which gives masses to the eight gluons
\cite{Ciminale:2006sm}. We therefore expect nine NGB due to
spontaneous breaking of the axial group $SU(3)$ and $U(1)_V$. The
eight bosons associated to $SU(3)$ have small masses while the
$U(1)_V$ boson (superfluid mode) is massless.

\subsection{Effective lagrangian of the superfluid
mode}\label{sec:U1} The superfluid mode is relevant for the
transport properties of the three flavor LOFF phase. Since its
lagrangian and parameters have not been derived before we give here
a brief account of this calculation, even though its derivation is
similar to that presented in \cite{Mannarelli:2007bs} for the
phonons associated to breaking of the rotational invariance.

The field $\phi$ is introduced as an external field by means of the
transformation $\psi\rightarrow U^\dagger \psi$ with $U =
\exp\left\{i\phi/f\right\}$~\cite{Eguchi:1976iz,Casalbuoni:2000na,Nardulli:2002ma}.
The quark lagrangian after the rotation reads
\begin{equation}
{\cal L} = \int\!\frac{d {\bm n}}{8\pi}\chi^\dagger_A
\left(\begin{array}{cc}
        V\cdot\ell \delta_{AB} + \delta\mu_{AB} & -\Xi_{BA}^\star \\
        -\Xi_{AB} & \tilde{V}\cdot\ell \delta_{AB} - \delta\mu_{AB}
      \end{array}
\right)\chi_B~,\label{eq:phi1}
\end{equation}
where
\begin{equation}
\Xi_{AB} =\Delta_I^\star({\bm r})\text{Tr}[\epsilon_I (F_A
U^\dagger)^T \epsilon_I F_B U^\dagger]~.\label{eq:phi2}
\end{equation}
From Eqs.~\eqref{eq:phi1} and~\eqref{eq:phi2} it is clear that the
field $\phi$ enters in the model as the phase of the order
parameter. The expansion of Eq.~\eqref{eq:phi1} at the lowest order
in $\phi$ gives rise to a three body and a four body interaction
lagrangians, namely
\begin{equation}
{\cal L}_{\chi\chi\pi} = \frac{2i\phi}{f}\int\!\frac{d {\bm
n}}{8\pi} \chi^\dagger_A\left(\begin{array}{cc}
        0 & -\Delta_{AB} \\
        \Delta^*_{AB} & 0
      \end{array}
\right)\chi_B~,\label{eq:Lagr3}
\end{equation}
\begin{equation}
{\cal L}_{\chi\chi\pi\pi} = -\frac{2\phi^2}{f^2}\int\!\frac{d {\bm
n}}{8\pi} \chi^\dagger_A\left(\begin{array}{cc}
        0 & -\Delta_{AB} \\
        -\Delta^*_{AB} & 0
      \end{array}
\right)\chi_B~.\label{eq:Lagr4}
\end{equation}
Next we integrate on the quark fields in the generating functional
of the model~\cite{Nardulli:2002ma},
\begin{equation}
W[\eta,\eta^\dagger] = \int {\cal D}\chi^\dagger{\cal D}\chi{\cal
D}\phi ~\exp \left\{i\!\int{\cal L} + {\cal L}_{\chi\chi\pi} + {\cal
L}_{\chi\chi\pi\pi} + \eta^\dagger\chi +
\chi^\dagger\eta\right\}~,\label{eq:partFunc}
\end{equation}
with ${\cal L}$ defined in Eq.~\eqref{eq:Lagr1}. The integration
procedure has been discussed in the literature both in the
homogeneous~\cite{Casalbuoni:2000na,Nardulli:2002ma,
Casalbuoni:2002st,Gatto:2007ja,Ruggieri:2007pi} and in inhomogeneous
cases~\cite{Mannarelli:2007bs}; in particular, in
Ref.~\cite{Mannarelli:2007bs} it was shown that in the LOFF phase,
where gapless quark excitations belong to the spectrum, one has to
introduce an infrared cutoff on the quark momenta, and integrate
over the fields with momenta greater than the cutoff; eventually one
sends the cutoff to zero. We apply the same procedure here. Once the
integration over the quark fields is performed one is left with the
effective action of $\phi$, ${\cal S} = {\cal S}_{s.e.} + {\cal
S}_{tad}$ with~\cite{Casalbuoni:2000na,Nardulli:2002ma,
Casalbuoni:2002st,Gatto:2007ja,Ruggieri:2007pi,Mannarelli:2007bs}:
\begin{equation}
{\cal S}_{s.e.} = +
\frac{i}{2}\left(\frac{2i\phi}{f}\right)^2\text{Tr}\left[S
\left(\begin{array}{cc}
        0 & -{\bm \Delta} \\
        {\bm \Delta}^* & 0
      \end{array}
\right)S \left(\begin{array}{cc}
        0 & -{\bm \Delta} \\
        {\bm \Delta}^* & 0
      \end{array}
\right)\right]~,\label{eq:SEsinglet}
\end{equation}
\begin{equation}
{\cal S}_{tad} = -
i\left(-\frac{2\phi^2}{f^2}\right)\text{Tr}\left[S
\left(\begin{array}{cc}
        0 & -{\bm \Delta} \\
        -{\bm \Delta}^* & 0
      \end{array}
\right)\right]~.\label{eq:TADsinglet}
\end{equation}
In the above equations $S$ is the quark propagator, and
${\bm\Delta}$ is the $9\times9$ gap matrix $\Delta_{AB}$; we use the
notation ${\cal S}_{tad}$ because the corresponding Feynman diagram
has a tadpole shape; on the other hand ${\cal S}_{s.e.}$ corresponds
to a self energy diagram. The trace is on space coordinates and over
all the internal degrees of freedom of the quarks, namely helicity,
color, flavor and Nambu-Gorkov.

In general the fermion propagator cannot be computed  exactly (an
exception is the single plane wave structure). Therefore one has to
employ some approximation; in this paper we use the Ginzburg-Landau
(GL) approximation,  based on an  expansion of $S$ in powers of
$\Delta/\delta\mu$. This approximation has been used
in~\cite{Mannarelli:2007bs} for the computation of the shear modulus
in the three flavor LOFF phase of QCD, evaluating the low energy
parameters in the phonon lagrangian at the second order in
$\Delta_I/\delta\mu_I$. We work here at the same order.

To begin with we consider the contribution \eqref{eq:SEsinglet}.
Evaluation of the traces gives
\begin{equation}
{\cal S}_{s.e.} = -i
\frac{2}{f^2}\sum_{I=2}^3\Delta_I^2~\sum_{a=1}^{P_I} \int\!\frac{d^4
k}{(2\pi)^4}~\phi(-k) \phi(k) ~{\cal P}_{I}^a(k_0,{\bm
k})~,\label{eq:se1}
\end{equation}
with $k = (k_0,{\bm k})$ and
\begin{eqnarray}
{\cal P}_{I}^a(k_0,{\bm k}) &=& -2\int\!\frac{d{\bm n}}{4\pi}
\int\!\frac{d^4 \ell}{(2\pi)^4} \left[\frac{1}{(\tilde
V\cdot\ell+\delta\mu-{\bm q}_I^a\cdot{\bm n})[V\cdot(\ell + k) + \delta\mu-{\bm q}_I^a\cdot{\bm n}] } \right. \nonumber \\
&& ~~~~~~~~~~~~~~~~~~ + \left.\frac{1}{( V\cdot\ell - \delta\mu-{\bm
q}_I^a\cdot{\bm n})[\tilde V\cdot(\ell + k)  -
\delta\mu-{\bm q}_I^a\cdot{\bm n}] } \right]\nonumber\\
&& ~~~~~~~~~+ \delta\mu \rightarrow -\delta\mu~.\label{eq:PaI}
\end{eqnarray}
On the other hand from \eqref{eq:TADsinglet} one gets
\begin{equation}
{\cal S}_{tad} = i
\frac{2}{f^2}\sum_{I=2}^3\Delta_I^2~\sum_{a=1}^{P_I}
\int\!\frac{d^4 k}{(2\pi)^4}~\phi(-k) \phi(k) ~{\cal P}_{I}^a(k_0
=0,{\bm k}=0)~.\label{eq:tad1}
\end{equation}
From Eqs.~\eqref{eq:se1} and~\eqref{eq:tad1} it is easily recognized
that one needs to evaluate ${\cal P}_{I}^a(k_0,{\bm k}) - {\cal
P}_{I}^a(0,{\bm 0})$, which is well-behaved in the ultraviolet. The
loop integral is evaluated in the usual way by Wick rotating to
imaginary energies $\ell_0 \rightarrow i\ell_4$. Since the integral
is convergent one can send the ultraviolet cutoff on
$\ell_\parallel$ to infinity, and perform the integral over
$\ell_\parallel$ by residues, followed by integration over $\ell_4$.
This calculation is similar to the one of
Refs.~\cite{Mannarelli:2007bs,Gatto:2007ja}, therefore we simply
quote here the final result of the lagrangian at small external
momenta, namely
\begin{equation}
{\cal L}(k) =  \frac{1}{2}\phi(-k)\left[k_0^2 {\cal I}_0 - k_i k_j
V_{ij}\right]\phi(k)~,\label{eq:Trallalla}
\end{equation}
where
\begin{eqnarray}
{\cal I}_0 &=& -\frac{\mu^2}{\pi^2 f^2}\sum_{I=2}^3 \Delta_I^2
P_I~\Re e\int\!\frac{d{\bm n}}{4\pi}\frac{1}{(\delta\mu - {\bm
q}_I^a\cdot {\bm n} + i0^+)^2}+ \delta\mu \rightarrow -\delta\mu~,
\\
V_{ij} &=& -\frac{\mu^2}{\pi^2 f^2}\sum_{I=2}^3 \Delta_I^2 P_I~\Re
e\int\!\frac{d{\bm n}}{4\pi}\frac{n_i n_j}{(\delta\mu - {\bm
q}_I^a\cdot {\bm n} + i0^+)^2}+ \delta\mu \rightarrow -\delta\mu~.
\end{eqnarray}

We specialize the results to the symmetric case $\Delta_2 = \Delta_3
\equiv \Delta$, $|{\bm q}_2^a| = |{\bm q}_3^a| \equiv q$, $P_2 = P_3
\equiv P$~\cite{Rajagopal:2006ig}. At the minimum one
has~\cite{Casalbuoni:2005zp} \be q=\eta\delta\mu~,~~~~~~~\eta
\approx 1.1997\ . \ee Requiring canonical normalization of the
Lagrangian in Eq.~\eqref{eq:Trallalla} implies ${\cal I}_0 = 1$,
that is
\begin{equation}
f^2 = \frac{4 P
\mu^2}{\pi^2}\frac{\Delta^2}{\delta\mu^2(\eta^2-1)}~.\label{eq:fSqPHI}
\end{equation}
The squared velocity tensor $V_{ij}$ can be computed following the
same steps as in~\cite{Gatto:2007ja}. For $P=1$ we find $V_{ij} =
[\text{diag}(0,0,1)]_{ij}$, where we have chosen the ${\bm q}_2$ and
${\bm q}_3$  along the positive $z$-axis. For the two cubic
structures corresponding to the values $P=4$ and $P=8$ we find
$V_{ij} = \delta_{ij}/3$, ie the velocity is isotropic and has the
value $1/\sqrt3$.

\subsection{Parameters of the $SU(3)_A$ Goldstone
bosons}\label{sec:R} The quark coupling to the octet of
pseudo-Goldstones can be introduced in a similar way
\cite{Eguchi:1976iz,Nardulli:2002ma}:
\begin{equation}
\psi_{\alpha i} \rightarrow \psi_{\alpha k}\left({\cal
U}^\dagger\right)_{ki}~,~~~~{\cal U} \equiv \exp
\displaystyle{\left\{i \frac{\pi_a
\lambda_a}{2F_a}\right\}}~,\label{eq:octet}
\end{equation}
where $a=1,...,8$, $\lambda_a$ are the Gell-Mann matrices,
normalized as $\text{Tr}\{\lambda_a \lambda_b\}=2 \delta_{ab}$~, and
$F_a$ are the decay constants relative to $\pi_a$. We remind that
Latin indices denote flavor, while Greek indices stand for color.

In order to describe the flavor excitations we promote the chemical
potential matrix to a spurion field with definite transformation
property under flavor transformation, namely $\mu\rightarrow L \mu
L^\dagger$ with $L\in SU(3)_L$ (analogously $\mu\rightarrow R \mu
R^\dagger$ with $R\in SU(3)_R$). This transformation leaves
invariant a chemical potential term under an $SU(3)_A$
transformation, and thus under ${\cal U}$ in Eq.~\eqref{eq:octet}.
The quark lagrangian after the rotation reads
\begin{equation}
{\cal L} = \int\!\frac{d {\bm n}}{8\pi}\chi^\dagger_A
\left(\begin{array}{cc}
        V\cdot\ell \delta_{AB} + \delta\mu_{AB} & -\Xi_{BA}^\star \\
        -\Xi_{AB} & \tilde{V}\cdot\ell \delta_{AB} - \delta\mu_{AB}
      \end{array}
\right)\chi_B~,
\end{equation}
where
\begin{equation}
\Xi_{AB} =\Delta_I^\star({\bm r})\text{Tr}[\epsilon_I (F_A {\cal
U}^\dagger)^T \epsilon_I F_B {\cal U}^\dagger]~.
\end{equation}

By expanding ${\cal U}$ in Eq.~\eqref{eq:octet} up to the second
order in the meson fields, we have a three body and a four body
interaction Lagrangians as in
Eqs.~\eqref{eq:Lagr3},~\eqref{eq:Lagr4} (the difference arising here
from the non trivial flavor structure of ${\cal U}$):
\begin{equation}
{\cal L}_{\chi\chi\pi}= +\frac{i\,\pi_a}{2F_a} \int\!\frac{d {\bm
n}}{8\pi}\chi^\dagger_A ({\cal G}_3)_{AB}^a\,
\chi_B~,\label{eq:octetLagr3}
\end{equation}
\begin{equation}
{\cal L}_{\chi\chi\pi\pi}= +\frac{\pi_a \pi_b}{8 F_a F_b}
\int\!\frac{d {\bm n}}{8\pi}\chi^\dagger_A ({\cal G}_4)_{AB}^{ab}\,
\chi_B~.\label{eq:octetLagr4}
\end{equation}
The expressions of ${\cal G}_3,{\cal G}_4$ are as follows:
\begin{equation}
{\cal G}_3=\left(\begin{array}{cc}
    0     & - (K^{3a}_{BA})^\star \\
       K^{3a}_{AB} & 0
      \end{array}
\right)~,\label{eq:octetLagrAPP3}
\end{equation}
\begin{equation}
{\cal G}_4=\left(\begin{array}{cc}
        0 & (K^{4 ab}_{BA})^\star  \\
      K^{4 ab}_{AB} & 0
      \end{array}
\right)~,\label{eq:octetLagrAPP4}
\end{equation}and the off-diagonal entries are defined as
\begin{eqnarray}
K^{3a}_{AB} &=& \Delta_I^\star({\bm r}) ~\text{Tr}[\epsilon_I
\lambda_a^T F_A^T \epsilon_I F_B
+\epsilon_I F_A^T \epsilon_I F_B \lambda_a ]~,\label{eq:K3}\\
K^{4ab}_{AB} &=& \Delta_I^\star({\bm r}) ~\text{Tr}[\epsilon_I
\lambda_a^T \lambda_b^T F_A^T \epsilon_I F_B + \epsilon_I F_A^T
\epsilon_I F_B \lambda_a \lambda_b + 2 \epsilon_I \lambda_a^T F_A^T
\epsilon_I F_B \lambda_b]~.\label{eq:K4}
\end{eqnarray}

From now on the steps leading to the effective lagrangian are
analogous to the previous case. However a complication arises from
the non-trivial flavor-color structure of the interaction vertices.
Integrating out the fermion fields in the generating functional one
gets the effective lagrangian in momentum space, ${\cal L}(p) =
{\cal L}_{s.e.}(p) + {\cal L}_{tad}$, with
\begin{eqnarray}
i{\cal L}_{tad} &=& +\left(\frac{\pi_a \pi_b}{8 F_a
F_b}\right)\frac{\mu^2}{4\pi^3 } \int\!\frac{d{\bm n}}{4\pi}\int
d^2\ell~\text{Tr}[ S(\ell)  {\cal
G}_4]~,\label{eq:tadpoleMom} \\
i {\cal L}_{s.e.}(p) &=& -
\frac{1}{2}\left(i\frac{\pi_a}{2F_a}\right)\left(i\frac{\pi_b}{2F_b}\right)
\frac{\mu^2}{4\pi^3}\int\!\frac{d{\bm n}}{4\pi}\int
d^2\ell~\text{Tr}[ S(\ell + p)  {\cal G}_3  S(\ell)  {\cal
G}_3]~.\label{eq:selfMom}
\end{eqnarray}
We have already kept into account the $L+R$ contribution, and the trace is on Nambu-Gorkov and
color-flavor indices. One gets
\begin{eqnarray}
{\cal L}_{s.e.}(p=0) &=&8\Delta_2^2 {\cal I}_2 ( \pi_1^2 + \pi_2^2)
+ 8\Delta_3^2 {\cal I}_2 ( \pi_1^4 + \pi_5^2) + 8(\Delta_2^2 +
\Delta_3^2)(-{\cal I}_1)(\pi_6^2 + \pi_7^2)
 \nonumber \\
&&~~+ 8\Delta_2^2(-{\cal I}_1)\pi_3^2 +
\frac{8}{3}\left(\Delta_2^2 + 4\Delta_3^2\right)(-{\cal I}_1)\pi_8^2 \nonumber\\
&&~~~+ \frac{16}{\sqrt{3}}\Delta_2^2 {\cal I}_1 \pi_3 \pi_8~,\label{eq:Pi1SE} \\
{\cal L}_{tad} &=&8\Delta_2^2 {\cal I}_1 ( \pi_1^2 + \pi_2^2) +
8\Delta_3^2 {\cal I}_1 ( \pi_1^4 + \pi_5^2) + 8(\Delta_2^2 +
\Delta_3^2) {\cal I}_1 (\pi_6^2 + \pi_7^2)
 \nonumber \\
&&~~+ 8\Delta_2^2 {\cal I}_1 \pi_3^2 + \frac{8}{3}\left(\Delta_2^2 +
4\Delta_3^2\right) {\cal I}_1 \pi_8^2 \nonumber \\
&&~~~- \frac{16}{\sqrt{3}}\Delta_2^2 {\cal I}_1 \pi_3 \pi_8
~,\label{eq:Pi1TAD}
\end{eqnarray}where
\begin{eqnarray}
{\cal I}_2 &=&\frac{-i }{8F^2}\frac{\mu^2}{4\pi^3}\sum_{a=1}^{P}
\int\!\frac{d{\bm n}}{4\pi}\int d^2\ell
\frac{(-1)}{(\ell_0+\ell_\parallel-2\delta\mu -{\bm q}^a\cdot{\bm n}
)(\ell_0
-\ell_\parallel-2\delta\mu-{\bm q}^a\cdot{\bm n})} \nonumber \\
&=&P\frac{\mu^2}{16\pi^2
F^2}\left(1-\frac{1}{\eta}\log\frac{2+\eta}{2-\eta} +
\frac{1}{2}\log\frac{\Lambda^2}{\delta\mu^2(4-\eta^2)}\right)\,
\label{eq:ooo}\cr {\cal I}_1 &=&\frac{-i
}{8F^2}\frac{\mu^2}{4\pi^3}\sum_{a=1}^{P} \int\!\frac{d{\bm
n}}{4\pi}\int d^2\ell
\frac{1}{(\ell_0+\ell_\parallel+2\delta\mu)(\ell_0
-\ell_\parallel-2{\bm q}^a\cdot{\bm n})} \nonumber \\
&=&-P\frac{\mu^2}{16\pi^2 F^2}\left(1-\frac{1}{2\eta}\log\frac{\eta
+ 1}{\eta - 1} + \frac{1}{2}\log\frac{\Lambda^2}{\delta\mu^2(\eta^2
- 1)}\right)~.\label{eq:ppp}
\end{eqnarray}In
these equations $\Lambda$ is an ultraviolet cutoff, needed since
both contributions are ultraviolet divergent; $\eta$ and $\delta\mu$
have been defined above.

We can take into account the effect of the light quark masses by
adding  an anti-gap term coupling two
antiquarks~\cite{Beane:2000ms}. According
to~\cite{Beane:2000ms,Schafer:2001za,Casalbuoni:2002st} we can write
the gap plus the anti-gap lagrangian in the form
\begin{eqnarray}
{\cal L} &=& \frac{\Delta^\star_I({\bm r})}{2}\psi^T_{i\alpha}C
\psi_{\beta j} \epsilon^{\alpha\beta I} \epsilon_{ijI} \nonumber
\\&- & \frac{\bar\Delta^\star_I({\bm
r})}{2}\frac{1}{4\mu^2}\psi^T_{k\alpha}C \psi_{\beta \ell}~M^T_{ik}
M_{\ell j}~\epsilon^{\alpha\beta I} \epsilon_{ijI} - L\rightarrow R
+ h.c.~,
\end{eqnarray}
where $\psi$ are left-handed and positive energy fields, and $M$ is
the quark mass matrix in flavor space. In the basis spanned by the
$F_A$ matrices the coupling of the Goldstones to the anti-gap is
obtained by rotating the quark field according to
Eq.~\eqref{eq:octet},
\begin{equation}
{\cal L} = \int\!\frac{d {\bm n}}{8\pi}\chi^\dagger_A
\left(\begin{array}{cc}
        0 & \Upsilon_{BA}^\star \\
        \Upsilon_{AB} & 0
      \end{array}
\right)\chi_B~,
\end{equation}
where
\begin{equation}
\Upsilon_{AB} =\frac{\Delta_I^\star({\bm
r})}{4\mu^2}\text{Tr}[\epsilon_I (F_A {\cal U}^\dagger M)^T
\epsilon_I F_B {\cal U}^\dagger M]~.
\end{equation}
Since the anti-gap gives contributions of the order of
$M_{u,d}M_s/\mu^2$, we treat it as an insertion. On the other hand
the corrections due to $\delta\mu=M_s^2/8\mu$ are computed exactly
in the GL approach, because in the kinematical region where this
expansion is valid, and the LOFF phase is favored, $\delta\mu$ is
rather large and  the small parameter is $\Delta/\delta\mu$.
Considering all the contributions we get the following results for
the boson masses (we put $\Delta_2=\Delta_3=\Delta$ and
$P_2=P_3=P$):
\begin{eqnarray}
&&m_{\pi^\pm}^2 = m_{K^\pm}^2 = c\,\frac{P\Delta^2\mu^2}{\pi^2 F^2}
~,\nonumber \\
&&m^2_{K^0} =m^2_{\bar K^0} = \frac{P\Delta\bar\Delta}{8\pi^2
F^2}M_u(M_d +
M_s)\log\frac{\mu^2}{\delta\mu^2(\eta^2 - 1)}~,\nonumber \\
&& m_{33}^2 = \frac{P\Delta\bar\Delta}{8\pi^2 F^2}M_u
M_s\log\frac{\mu^2}{\delta\mu^2(\eta^2 - 1)}~,\nonumber
\\
&& m_{88}^2 = \frac{P\Delta\bar\Delta}{24\pi^2 F^2}(M_u M_s + 4 M_u
M_d)\log\frac{\mu^2}{\delta\mu^2(\eta^2 - 1)}~,\nonumber
\\
&& m_{38}^2 = -\frac{P\Delta\bar\Delta}{8\sqrt{3}\pi^2 F^2}M_u
M_s\log\frac{\mu^2}{\delta\mu^2(\eta^2 - 1)}~. \label{eq:Pi1tot}
\end{eqnarray}The numerical value of $c$ is $c\simeq 1.03$, obtained
using the numerical value of $\eta$ in the expressions of ${\cal
I}_1$ and ${\cal I}_2$; moreover we have introduced the fields
$\pi^\pm = (\pi_1 \mp i \pi_2)/\sqrt{2}$, $K^\pm = (\pi_4 \mp i
\pi_5)/\sqrt{2}$, $K^0/{\bar K}^0 = (\pi_6 \mp i \pi_7)/\sqrt{2}$.
In the above mass formulae we have neglected light quark mass
effects in the case of the charged pions, since they are suppressed
in comparison  to the leading order result, see
Eqs.~\eqref{eq:Pi1SE} and \eqref{eq:Pi1TAD}.

Several comments are in order. First, we find $m_{K^0} \neq
m_{K^+}$. This is easily explained by noticing that the $SU(2)$
isospin symmetry in the light quark sector is explicitly broken by
$\mu_u \neq \mu_d$. As a matter of fact if in the quark loops we
would put $\mu_u = \mu_d$, thus restoring isospin symmetry, then we
would obtain $m_{K^0} = m_{K^+}$. Second, in the limit $\Delta_2 =
\Delta_3$ we find $m_{\pi^+} = m_{K^+}$. From the diagrammatic point
of view this equality is explained in the following way. The tadpole
diagram of $K^+$ is obtained from the $\pi^+$ tadpole by replacing
$\Delta_2 \rightarrow \Delta_3$ and $\mu_s \rightarrow \mu_d$, which
is equivalent to  $\delta\mu \rightarrow -\delta\mu$, Since the
tadpole diagram does not depend on the sign of $\delta\mu$, then the
equality of the two diagrams follows. Similarly, the $K^+$
self-energy diagram is obtained from the $\pi^+$ one by replacing
$\Delta_2 \rightarrow \Delta_3$: also in this case the equality is
achieved since $\Delta_2 = \Delta_3$.

The mass formulas of the three flavor LOFF phase depend on the light
quark masses similarly to the CFL phase, see for
example~\cite{Beane:2000ms}. The differences arise because here we
have adopted the expansion of the quark propagator in powers of
$\Delta/\delta\mu$, resulting in a different argument in the
logarithm (in the CFL one has $\log(\mu/\Delta)$ instead of
$\log(\mu/\delta\mu)$); moreover, in the LOFF phase  terms
containing the product $M_d M_s$ are not present since they are
proportional to $\Delta_1$, which is zero in our
approximation.

Finally, we find that all the squared masses of the pseudo-Goldstone
modes are positive. Therefore, at least to the order ${\cal
O}(\Delta^2/\delta\mu^2)$, there is not meson condensation in the
three flavor LOFF phase.

By the same procedure we can compute the decay constants of the
octet. Expanding the self-energy lagrangian up to the second order
in the external momentum of the bosons, and requiring the lagrangian
to be canonically normalized, we find
\begin{eqnarray}
 F_{\pi^\pm}^2 &=&  \frac{P \Delta^2\,\mu^2}{8 \pi^2
\delta\mu^2 (4-\eta^2)} ~,\nonumber  \\
 F_{K^\pm}^2 &=& \frac{P \Delta^2\,\mu^2}{8 \pi^2 \delta\mu^2
(4-\eta^2)}~, \nonumber  \\
 F_{K^0}^2 &=& F_{\bar K^0}^2 =\frac{2P \,\mu^2\, \Delta^2} {8 \pi^2
\delta\mu^2
(\eta^2-1)} ~,\nonumber \\
 F_{33}^2 &=& \frac{P \Delta^2\,\mu^2}{8 \pi^2 \delta\mu^2
(\eta^2-1)} ~,\nonumber \\
 F_{88}^2 &=& \frac{5P \,\mu^2\,\Delta^2}{24 \pi^2
\delta\mu^2(\eta^2-1)} ~,\nonumber \\
 F_{38}^2 &=& \frac{P \,\mu^2\,\Delta^2}{8 \sqrt{3} \pi^2
\delta\mu^2(\eta^2-1)}~.\label{eq:YYY}
\end{eqnarray}

\section{Conclusions\label{concl}}
In this paper we have computed the parameters of the low energy
effective action of the meson excitations (Goldstone modes) in the
cubic structures of the three flavor LOFF phase of QCD. Since in the
LOFF state we are not able to write exactly the full quark
propagator we use an approximation, obtained by the expansion in
$\Delta/\delta\mu$, to the order $\Delta^2/\delta\mu^2$. We consider
the mode corresponding to the breaking of $U(1)_V$ (superfluid mode)
and the octet of  scalar fields related to $SU(3)_A$. The motivation
of this work was twofold. First, since the superfluid mode is
massless even in presence of finite quark masses, it is relevant for
the low energy dynamics of the LOFF phase. Thus, if LOFF quark
matter is present in the core of a compact star, the superfluid mode
should have a role for quark transport properties. Therefore the
computation of its low energy parameter can be of interest for
applications. Preliminary investigations of the astrophysical
effects of the LOFF state are in~\cite{Anglani:2006br}.

The second motivation was to study the possibility of pseudoscalar
field condensation in the octet sector, in order to see if,
similarly to the CFL phase, one has such effect also the three
flavor LOFF phase. To this end we have evaluated the octet mass
matrix.  We have found that the squared mass tensor is positive
defined, hence excluding the possibility of scalar condensation (at
least to the order $\Delta^2/\delta\mu^2$). Since all the masses are
non-vanishing, the octet is not expected to play an important role
in the low energy dynamics. As a conclusion, the low energy
effective theory for the three flavor LOFF phase should include the
gapless fermions, the superfluid mode and the phonon fields related
to the deformations of the crystal lattice~\cite{Mannarelli:2007bs}.

\acknowledgments We thank M.~Mannarelli and R.~Sharma for
enlightening correspondence. We would like to thank the CERN Theory
Division for the kind hospitality.

\end{document}